\documentstyle[aps,epsf]{revtex}
\tighten

\begin{document}

\title{Coherent Random Lasing and "Almost Localized" Photon Modes}

\author{V. M. Apalkov$^1$ \and  M. E. Raikh$^1$ \and B. Shapiro$^2$}
\address{
$^1$Department of Physics, University of Utah, Salt Lake City,
UT 84112, USA \\
$^2$Department of Physics, Technion-Israel Institute of
Technology, Haifa 32000, Israel}
\maketitle

\begin{abstract}
 A pulse of light, injected into a weakly disordered dielectric medium,
typically, will leave its initial location in a short time, by
diffusion.  However, due to some rare configurations of disorder,
there is a possibility of formation of high quality resonators which
can trap light for a long time.  We present a rather detailed,
quantitative study of such random resonators and of the "almost
localized" states that they can support. After presenting a brief
review of the earlier work on the subject, we concentrate on a
detailed computation of the "prefactor": knowledge of the latter is
crucial for varifying the viability of the random rasonators and their
areal density.  Both short range disorder (white noise) and correlated
disorder are studied, and the important effect of the correlation
radius, $R_c$, on the probability of formation of resonators with a
given quality factor $Q$ is discussed. The random resonators are
"self-formed", in the sense that no sharp features (like Mie
scatterers or other "resonant entities") are introduced: our model is
a featureless dielectric medium with fluctuating dielectric constant.
We point out the relevance of the random resonators to the recently
discovered phenomenon of coherent "random" lasing and review
 the existing work on that subject. We emphasize,
however, that the random resonators exist already in the {\em passive}
medium: gain is only needed to "make them visible".
\end{abstract}

\section{Introduction}
 There exists a formal analogy between the Schr\"{o}dinger
equation describing electron motion in a random potential,
 and the scalar
wave equation for light propagation in a medium with fluctuating
dielectric constant. Consider,  for concreteness, a two-dimensional
geometry (quantum well for an electron and a thin film for light).
Then both  equations can be presented in the form
\begin{equation}
\Delta _{\bbox{\rho }} \psi + \left[ k^2 - U(\bbox{\rho }) \right]
\psi = 0, \label{Eq1}
\end{equation}
where $\bbox{\rho}$ is the in-plane coordinate.
For an electron with mass $m$ and energy $E$, moving in a random
potential
$V(\bbox{\rho })$, the parameters
 $k^2$ and $U(\bbox{\rho })$ are
\begin{equation}
k^2 = 2m E~,~~~ U(\bbox{\rho }) =  2m V(\bbox{\rho }).
\end{equation}
For a light wave with frequency $\omega$,  traveling in a medium with
dielectric constant $\epsilon$,
corresponding expressions for $k^2$ and $U(\bbox{\rho })$ take the form
\begin{equation}
k^2 = \epsilon \left( \frac{\omega }{c} \right)^2~, ~~~~
     U(\bbox{\rho }) = -\delta \epsilon (\bbox{\rho })
    \left( \frac{\omega }{c} \right)^2,
\label{light}
\end{equation}
where $\delta \epsilon (\bbox{\rho })$ is the fluctuating part of the
dielectric constant.

The analogy between electrons and light is incomplete. For one
thing, the "scattering potential" for light depends on frequency
and it vanishes in the limit of zero frequency. Moreover, in a
dielectric medium (positive $\epsilon$) $k^2$ must be positive
which implies that there can be no bound states for light, as long
as $\delta \epsilon (\bbox{\rho })$ is assumed to vanish outside
some finite region in space. Therefore the simple notion of a
binding potential well does not exist for light: for instance, a
dielectric sphere embedded in a uniform medium cannot bind
photons, regardless of whether its dielectric constant is larger
or smaller than that of the surrounding medium. Thus, unlike
electrons, photons will always escape from a dielectric medium
into the surrounding air. However, under appropriate conditions,
they can be trapped within the sample for a long time. The main
purpose of this paper is to discuss trapping of light in a weakly
disordered dielectric film.

 The random term in
Eq. (\ref{Eq1}) is zero on average; its statistical properties are
described by the r.m.s. value, $U_0$, and the correlator
${\mbox{\large $K$}(\bbox{\rho_1}- \! \bbox{\rho_2})}$
\begin{equation}\label{correlator}
\left\langle U(\bbox{\rho }) \right\rangle = 0 ~,~~~
   \left\langle U(\bbox{\rho _1 }) U(\bbox{\rho _2})
 \right\rangle = U_0^2 \mbox{\large $K$}(\bbox{\rho _1 }-
\bbox{\rho _2})~, ~~~~\mbox{\large $K$}(0) =1.
\end{equation}
It is known \cite{abrahams} that in two dimensions
the nature of solutions
of Eq. (\ref{Eq1}) is governed by the dimensionless conductance $kl$,
where $l$ is the transport mean free path. Using the golden rule,
the product $kl$ can be expressed through the correlator,
$\mbox{\large $K$}$,
as follows
\begin{eqnarray}
 (kl)^{-1}  & = & \frac{U_0^2}{4 \pi k^4}  \int dq~ d\phi ~ q^3
  \delta (q^2 + 2 kq \cos \phi )
  \int d^{~\!\!2} \! \rho
~\mbox{\large $K$}(\rho ) \exp(i q \rho )  =  \nonumber \\
   & = & \frac{2 U_0^2}{k^2}
 \int _0 ^{\pi /2}d \alpha ~ \sin^2 \alpha  \int _0 ^{\infty }
  d \rho ~ \rho ~
\mbox{\large $K$}(\rho )
     ~J_0(2k\rho \sin \alpha ),
\label{kl1}
\end{eqnarray}
where $J_0$ is the Bessel function of zero order.
The above integral can be evaluated
analytically if we choose
a gaussian form for the correlator
$\mbox{\large $K$}(\rho ) = \exp (-\rho^2 /R_c^2)$.
Substituting this form into (\ref{kl1}), we obtain
\begin{equation}
(kl)^{-1} = \pi \left(\frac{U_0 R_c}{2k}\right)^2 ~\!\!
\mbox{\large $F$} \left( \frac{k^2R_c^2}{2} \right),
\label{kl1_2}
\end{equation}
where the dimensionless function $\mbox{\large $F$}$ is defined as
\begin{equation}
\mbox{\large $F$}(x) = e^{-x} \mbox{\Large $[$}
    I_0(x) -  I_1(x)  \mbox{\Large $]$}.
\end{equation}
Here $I_0$ and $I_1$ are the modified Bessel functions of  zero
and  first order, respectively.

As a simple example of a realistic  two-dimensional disorder,
consider a system of disks with random positions of the centers and with
fluctuating radii described by a normalized
distribution function, $\Phi (R)$. Within a disk, we have
$U(\bbox{\rho})=U_d$, whereas outside the disk
 $U(\bbox{\rho})=0$.
If the filling fraction, $f$, is low,
 $f \ll 1$,  the positions of the centers of the disks are
uncorrelated. Then  $\mbox{\large $K$}(\rho )$ is
determined by the overlap between
two circles of the same radius
with centers shifted by $\rho $. A straightforward calculation yields
\begin{eqnarray}
& &  \left\langle U(\bbox{\rho _1 }) U(\bbox{\rho _1}+\bbox{\rho })
 \right\rangle
=
 \frac{2U_d^2f
}{\pi}\!\! \int
     dR ~  \mbox{\large $\theta $}\!\! \left(R -\frac{\rho}{2}\right)
\Phi(R)   \nonumber \\
& & ~~~~~~~~~~~~~~~~~~~~~~~~~~~~  \times  \left\{
     \arcsin \sqrt{1- \! \left(\frac{\rho}{2R}\right)^2} -\!
\frac{\rho }{2R}
   \sqrt{1-\! \left(\frac{\rho}{2R}\right)^2}
     \right\},
\label{intR}
\end{eqnarray}
where $ \mbox{\large $\theta $}(x)$ is the step-function.
The actual distribution function, $\Phi (R)$, is governed
by various technological factors. However, it is reasonable to
assume that small values of $R$ are strongly unlikely, and that
$\Phi (R)$ falls off abruptly at large $R$. Consider as an example
the distribution $\Phi(x)=c^3 x^4 \exp(- cx^2)$, where  $x =
R/\overline{R}$ ($\overline{R}$ is the average radius) and
$c=64/9\pi$. The distribution is designed so as to yield the value
$50\%$ for the relative spread in $R$, i.e., $\langle\vert R-
\overline{R}\vert\rangle /\overline{R}=0.5$.
The result of calculation of the integral (\ref{intR}) using
this distribution is shown in Fig.~1 together with
its gaussian fit, which yields $U_0^2=0.86U_d^2f$ and
$\mbox{\large $K $}(\rho )=\exp \left(-3.4 \rho^2/\overline{R}^{~\!2}
\right)$. From this fit we conclude that the gaussian
form of the correlator,
$ \mbox{\large $K $}(\rho )$,
corresponds to a quite generic distribution of $R$.

\begin{figure}
\centerline{
\epsfxsize=3.0in
\epsfbox{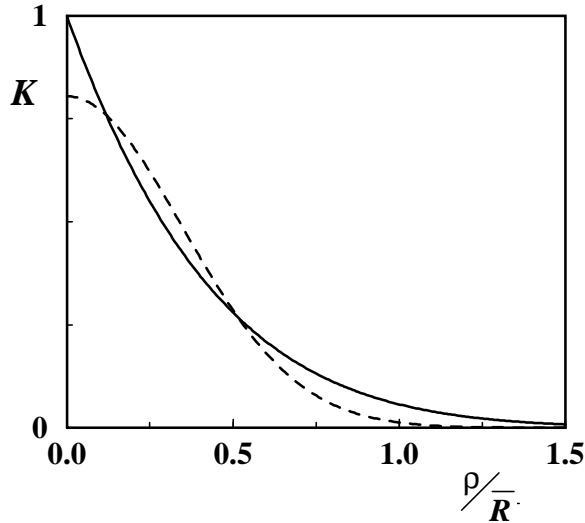}}
\vspace*{0.07in}
\caption[]{
Correlation function (solid line) for the model of
randomly distributed disks
$U(\bbox{\rho }) = U_d \theta ( R -\rho) $ with fluctuating radii
is shown together with its gaussian fit (dashed line).}
\label{eps1}
\end{figure}

Equation (\ref{kl1_2}) provides the explanation why the criterion for
strong localization, $kl < 1$, can be easily satisfied for electrons,
but very hard to achieve in the case of photons.
Indeed, for electrons Eq. (\ref{kl1_2}) can be presented as
\begin{equation}
\label{kl2}
(k l)^{-1} = \left( \frac{\pi m }{2E}\right) V_0^2 R_c^2
~\mbox{\large $F$} \left( mE R_c^2\right),
\end{equation}
while for photons Eq. (\ref{kl1_2}) takes the form
\begin{equation}\label{kl3}
(k l)^{-1} = \left( \frac{\pi }{4 \epsilon }\right)
   \left( \frac{\omega }{c} \right)^2 \Delta^2 R_c^2
~\mbox{\large $F$} \left( \frac{\epsilon \omega ^2R_c^2}{2c^2} \right),
\end{equation}
where $V_0$ and $\Delta $  are the r.m.s. fluctuations of
the potential and dielectric constant, respectively.
It is seen from Eq. (\ref{kl2}) that, since the function $F$
is always smaller than one, the electrons are strongly localized
in the {\em low-energy}  domain $E < E_c$. The value of
 $E_c$ is different for
the short-range, $R_c \ll \left(mV_0\right)^{-1/2}$,
and for the smooth,
$R_c \gg \left(mV_0\right)^{-1/2}$, potentials. For short-range
potential we can use the asymptotics
$F(x)\mbox{\Large $\vert$}_{x \ll 1}\approx 1$, which leads us to the
estimate $E_c \sim mV_0^2R_c^2$. Hence,
$mE_cR_c^2 \sim \left(mV_0R_c^2\right)^2 \ll 1$, so that the argument of
$F$ is  indeed small at $E=E_c$. In the case of the smooth potential,
the condition $mV_0R_c^2 \gg 1$ suggests that
the semiclassical description applies, so that $E_c \sim V_0$.
Calculation of the mean free path based on the golden rule is inadequate
in this case. To see this, note that at $E \sim V_0$, the argument
of $F$ in Eq. (\ref{kl2}) is large, so that we can use the asymptotics
$F(x)\mbox{\Large $\vert$}_{x \gg 1}\propto x^{-3/2}$. Then
Eq. (\ref{kl2}) yields $l \sim R_c$ for $E \sim V_0$. Thus, for smaller
$E$, namely  $E < V_0$, we have $l < R_c$. The latter
relation indicates
the failure of the perturbation theory for $E < V_0$.

Now let us perform the similar analysis for photons.  Using the
 small-$x$ and large-$x$ asymptotics of the function $F$, we obtain
 from Eq. (\ref{kl3})
\begin{equation}
\label{kl4}
 kl |_{k_0R_c \ll 1} =  \frac{4 \epsilon  }{\pi (k_0 R_c\Delta )^2}  ~,~
kl |_{k_0R_c \gg 1}=  \frac{4\epsilon ^{5/2} k_0R_c}{\pi ^{1/2}
\Delta ^2 },
\end{equation}
where $k_0 =\omega/c$. Eq. (\ref{kl4}) suggests that when the r.m.s.
fluctuation $\Delta$ is weak, $\Delta \ll \epsilon$, we have $kl \gg 1$
{\em both} in the low-frequency ($k_0R_c \ll 1$) and in the
high-frequency ($k_0R_c \gg 1$) domains.
This peculiar result is due to the already mentioned frequency
dependence of the "optical potential".

 The above analysis, however, does not rule out the possibility
of light  localization in the {\em strongly scattering} media.
In fact, first experimental indications of localization effects
for microwaves
were reported more than a decade ago  \cite{garcia91,genack91}.
In these experiments  localization  was inferred from the
measurements of various transmission characteristics of the
microwaves  through the tube filled with a random mixture of
aluminum and Teflon spheres.
For optical frequencies \cite{wiersma97,scheffold99,wiersma99}, the
strongly scattering medium used in first experiments, aimed at
light localization, was a semiconductor ({\em GaAs}) powder.
Measurements of the transmission vs. the sample thickness were
complimented with measurements of the coherent backscattering
(CBS) cone \cite{albada85,wolf85}. In the latter measurements
localization manifests itself through the rounding of the CBS cone
by limiting the maximal length of coherent path \cite{akkermans88}.

The early reports of the observation of  wave localization
both for microwaves and for
light [2-4] were inconclusive because
of the possibility that the results were affected by  absorption.
To get rid of this ambiguity, in the later experiments the CBS
mesurements \cite{schuurmans99} were performed on the macroporous
{\em GaP} networks, which scatter light stronger than a
powder \cite{wiersma97}. This allowed the authors to rule out the
absorption or the finite sample size as a source of rounding of
the CBS cone. For
microwaves [11-13], the recent
progress in detecting localization is due to a novel approach to
the analysis of the transmission data based on analysis of the
relative size of the transmission fluctuations.  This approach
permits one to detect localization even in the presence of absorption.

In the weakly scattering {\em active} medium the propagation of
light remains diffusive.
However, the interplay of the diffusion and the gain-induced
amplification can be
very nontrivial. Namely, this interplay can give rise to the
{\em incoherent} random lasing,
predicted by V.~I.~Letokhov  \cite{letokhov}.
As it was pointed out in Ref.  \cite{letokhov}, there is a close
 analogy between multiplication of neutrons in course of the chain
reaction and photons in the amplifying disordered medium
(photonic bomb). Upon
first experimental
observation of incoherent random lasing \cite{Law94}, it was subsequently
reproduced for various realizations of the gain media and different
types of disorder. Comparison of  theoretical [16-21]
 and experimental results [15,22-42] has confirmed that the
diffusion theory, which neglects the
interference effects,  is  quite sufficient for the
 description of {\em incoherent} lasing.
 Except for studies on powder grains of laser crystal materials
 [37-40] and $\pi$-conjugated polymer films
 \cite{Hide96,Denton97}, the majority of experiments [15,22-36]
have used dye solutions as amplifying media. Colloidal particles
suspended in a solution served as random scatterers. These scatterers
are responsible for {\em nonresonant feedback} required for incoherent
lasing. The essence of this feedback is that the light
amplification length,
$\tilde l_a$  (in the absence of  disorder), is significantly shortened
$[$to  $\sim (l\tilde l_{a})^{1/2}]$ in the disordered
medium,
when the light propagation is diffusive. The  threshold condition for
incoherent lasing corresponds to the gain magnitude at which
$(l\tilde l_{a})^{1/2}$ becomes of the order of the sample size.

In contrast to incoherent lasing, the recent discovery of
{\em coherent} random lasing [43-46] adds a new
dimension to the physics of light propagation in disordered media.
Coherent random lasing emerges as the degree of disorder increases, so
that the mean free path, $l$, becomes progressively smaller.
The fact that the light, emitted from a disordered sample, is truly
coherent, which does not necessarily follow  \cite{Wiersma2000,Wiersma2001}
from drastic narrowing of the emission spectrum, observed in
Refs. [43-46],
 was later convincingly demonstrated in photon statistics
 experiments  \cite{cao2001,Polson01}.
In the absence of mirrors, it is evident that, in order to support
the coherent lasing, the disordered medium {\em itself} should
assume their role. The latter is feasible only due to the {\em
interference effects}, that are not captured within the diffusion
picture.  More quantitatively, in order for the random medium to
play the role of a Fabry-Perot resonator, it is necessary that
certain eigenfunctions of Eq.~(\ref{Eq1}) were either completely
localized or {\em almost} localized. Almost localized solution can
be, roughly, envisaged as a very high {\em local} maximum of the
extended eigenfunction $\psi(\bbox{\rho })$ of Eq.~({\ref{Eq1}).
If this maximum is viewed as a {\em core} of $\psi(\bbox{\rho })$,
then the delocalized {\em tail} (see Fig. 2) can be viewed as a
source of {\em leakage}. In other words, the core itself, being
not an exact eigenfunction, can be viewed as a solution of
Eq.~(\ref{Eq1}) corresponding to a {\em complex} eigenvalue
$\mbox{Im}~\! k^2 \neq 0$.  Then the weak leakage translates into
a small value of the imaginary part of $k^2$. Inverse of this
imaginary part determines the {\em lifetime} of the core. The
higher is the local maximum of $\psi(\bbox{\rho })$, the longer is
the lifetime. Making link to the Fabry-Perot resonator, the ratio
$ k^2 /\mbox{Im}~\! k^2 $ can be identified with a quality factor,
$Q$. Thus, from the perspective of the coherent random lasing, the
right question to be asked is: how high are the attainable quality
factors of the almost localized solutions of Eq.~(\ref{Eq1}) at a
given prameters of disorder, i.e., magnitude, $\Delta $, and
correlation radius, $R_c$.

\begin{figure}
\centerline{
\epsfxsize=3.4in
\epsfbox{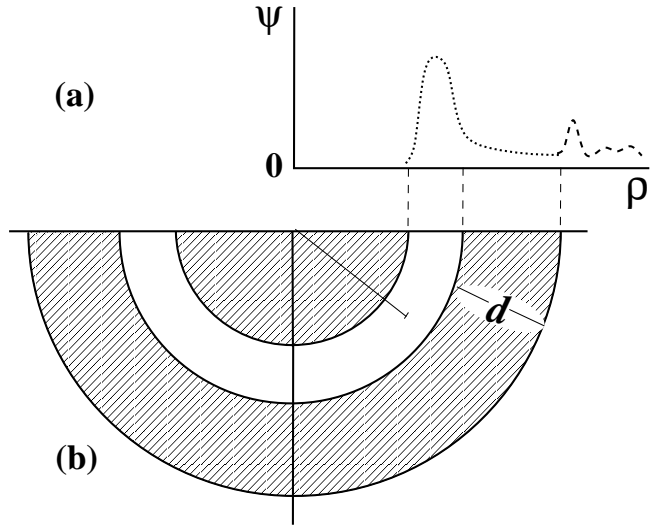}}
\vspace*{0.07in}
\caption[]{(a) Spatial distribution of  the wave function of the
anomalously localized solution, $\psi (\rho )$,
 of Eq.~(\ref{Eq1}) is shown schematically.
(b) Spatial distribution of the dielectric constant,  $\epsilon (\rho )$,
 corresponding to the trap, responsible for the solution $\psi (\rho )$;
$\epsilon (\rho ) = \epsilon $ outside the blank region.
Only the lower half of the trap is shown.}
\label{eps3}
\end{figure}

We address this question in the present paper.  However, prior to
discussing the almost localized
solutions of Eq. (\ref{Eq1}), a subtle point must be clarified.
Namely, Eq. (\ref{Eq1}) with the ``potential'' being real,
describes propagation of light in a {\em passive} medium.  It
might be argued that, in the presence of gain, which is required
for lasing, the spatial structure of the solutions of Eq.
(\ref{Eq1}) undergoes a drastic change, so that the almost
localized modes of a passive random medium and actual {\em lasing}
modes have little in common. In fact, it is well known both from
theory \cite{zyuzin94} and from CBS experiments  \cite{elongated}
that the {\em diffusive} trajectories of light within a random
medium get {\em elongated} in the presence of the gain. With
regard to coherent random lasing, it was initially
claimed \cite{cao00} that the gain {\em facilitates localization}
of light. This claim was even supported by the numerical
simulations \cite{cao00}. However, in the later
theoretical [54-56] and
experimental \cite{cao02} papers it was explicitly stated that,
similarly to the conventional lasers with Fabry-Perot resonators,
the gain only {\em reveals} high-$Q$ solutions of Eq. (\ref{Eq1})
existing in the passive medium. Thus, in the present paper, we
will focus exclusively on the passive disordered media. The
question about the likelihood of formation of the disorder-induced
resonators is central to the understanding of the coherent random
lasing. This question is at the core of the ongoing  in-depth
experimental studies [57-65].
 Except for
Ref.~ \cite{apalkov02}, theoretical papers on random
lasing [56,67-71] do not address this question.

Let us finally mention that a similar question of trapping
electrons, for a long time, in a weakly disordered conductor was
addressed long ago in the context of electronic
transport [72-75]. However,
the "almost localized" states discussed in
 \cite{apalkov02,karpov93,karpov'93} and,
further, in this paper are quite different from the "prelocalized"
states studied in [72-75].
Wave functions of the "almost localized" states are confined
primarily to a small ring (a high-$Q$ resonator). Moreover, these
states are extremely sensitive to the value of the correlation radius,
$R_c$, of the disordered potential. We shall see below that, for a
given parameter $kl>1$, the likelihood of high-$Q$ "almost
localized" modes is very low for the  short-range disorder but
sharply increases with $R_c$. Such features are absent for the
"prelocalized" states [72-75], with their
comparatively large spatial extent.

\section{Likelihood of Random Resonators in a Disordered Film}

\subsection{Intuitive Scenario of Random Cavities Based on
 Recurrent Scattering}

The first experimental work on coherent random lasing by Cao {\em
et. al.} \cite{cao98} was carried out on thin (with a width $\sim
2\pi/k_0$) zinc oxide (ZnO) polycrystalline films. Laser action
manifested itself through a drastic narrowing of the emission spectrum
when the optical excitation power exceeded a certain threshold.  The
authors of Ref.  \cite{cao98} realized at the time that coherent lasing
requires a resonator (cavity). They conjectured that, in the absence
of traditional well-defined resonator (as in a semiconductor laser),
the cavities in polycrystalline films are ``self-formed'' due to
strong optical scattering.  For a microscopic scenario of the
formation of such cavities, they alluded to the remark made in
Ref.  \cite{Wiersma96} that closed-loop paths of light can serve as
``random ring cavities''. The importance of these closed-loop paths
(recurrent scattering events) was pointed out earlier \cite{wiersma95},
when they were invoked for the explanation of the magnitude of the CBS
albedo.  With regard to CBS, the effect of recurrent scattering
events, e.g., the events in which
the {\em first} scattering
(of the incident wave)
and the {\em last} scattering (of the outgoing wave)
are provided by the {\em same} scatterer,
is that these events {\em do not} contribute to the CBS, thus reducing
the albedo in the backscatering direction.

The fact that recurrent events show up in CBS certainly does not allow
to automatically identify closed-loop paths with resonator cavities.
Therefore, the feasibility of the scenario of random cavities, adopted
in Refs.  \cite{cao98,cao99,cao99'',cao99'} for interpretation of
experimental results, was later put in question \cite{thesis}. The
arguments against this scenario were the following. Since in each
scattering act most of the energy gets scattered out of the loop, an
unrealistically high gain would be required to achieve the lasing
threshold condition for such a loop. Also the loops of scatterers are
likely to generate a broad frequency spectrum rather than isolated
resonances.

Certainly the picture of random cavities representing a certain
spatial arrangement of {\em isolated} scatterers is too naive.  This,
however, does not rule out the entire concept of disorder-induced
resonators. Although  sparse, the disorder configurations
that trap the light
for long enough time can be found in a sample of a large enough size, and
a single such configuration is already sufficient for lasing to
occur.  Therefore, under the condition $kl \gg 1$, which implies that
{\em overall} scattering is weak, the conclusion about the relevance of
random cavities can be drawn only upon the
{\em calculation} of their likelihood.
This calculation is decribed in the rest of this Section first on qualitative,
and then on quantitative levels.

Making link to the discussion in the Introduction: by pursuing the
random cavity scenario Cao {\em et. al.}  have intuitively
arrived to the concept of
prelocalized states, that was introduced in transport more than a decade
ago. The fundamental difference between the prelocalized
states and recurrent
events is that recurrent events emerge in the higher order [in
parameter $kl^{-1}$] of the {\em perturbation theory} (they
correspond to the specific type of diagrams called
Hikami-boxes \cite{hikami81}), whereas the formation of prelocalized
states is a genuinely {\em nonperturbative} effect. Already in the
first analytical approach to the problem of prelocalized
states \cite{altshuler86}, it was demonstrated that, in order to
capture an anomalously slow tail of the conductance relaxation (which
is due to trapping), {\em all} the orders of the perturbation theory
must be taken into account.

\subsection{Optimal Fluctuation Approach to the Problem}

\subsubsection{Qualitative Discussion.}

Similarly to the treatment in
 \cite{cao98,cao00,cao99'} we restrict our consideration to
the two-dimensional case (a disordered film).
Regarding the geometry of a random resonator, we adopt
the idea  proposed by Karpov
 for trapping   acoustic waves  \cite{karpov93} and  electrons
 \cite{karpov'93} in three dimensions.

Suppose that within a certain stripe the effective in-plane dielectric constant
of a film is  enhanced  by some small value $\epsilon _1 \ll \epsilon $.
Then such a stripe can play a role
of a {\em waveguide}, i.e., it can capture a transverse mode, as it is
illustrated in Fig.~3. There is no threshold for such a waveguiding, which
means that the transverse mode will be captured even if the width of the stripe
is small. Now, in order to form a resonator, one has to roll the stripe into a ring.
Upon this procedure,  the mode propagating along the waveguide
 transforms into a whispering-gallery mode of a ring. An immediate consequence
of the curving of the waveguide is emergence of the evanescent
leakage - the optical analog of the under-the-barrier tunneling in
quantum mechanics  (see Fig.~2). This leakage is responsible for a
finite lifetime of the whispering-gallery mode. Thus we have
specified the structure of the  weakly decaying solutions of Eq.
(\ref{Eq1}), discussed in the Introduction. Namely, the mode of
the waveguide
 plays the role of the {\em core}, while delocalized tail (see Fig.~2) reflects the
evanescent leakage.
Due to the azimuthal symmetry, the modes of the resonator
are characterized by the angular momentum,
$m$. Denote by $\mbox{\bf  N}_m(kl,Q)$ the areal density of resonators
with quality factor $Q$ in the film with a transport mean free path
$l$. Obviously, in the diffusive
regime, $kl > 1$, the density $\mbox{\bf  N}_m(kl,Q)$ is exponentially
small for $Q \gg 1$. In this domain $\mbox{\bf  N}_m(kl,Q)$ can be
 presented as
\begin{equation}\label{concentration}
\mbox{\bf  N}_m(kl,Q) \propto
   e^{-S_m(kl, Q)} ~.
\end{equation}
Let us first give a qualitative estimate for $S_m$, which reveals
its sensitivity to the strength and the range of the disorder
($\Delta ^2$ and  $R_c$). Since $m$ is the number of wave lengths
along the ring, its radius is $\rho _0 = m/k $. The ring waveguide
can support a weakly decaying mode only if its width $w$ satisfies
the condition $w(\epsilon _1/\epsilon )^{1/2} > k^{-1}$.  A
straightforward estimate for the decay time due to evanescent
leakage, i.e., the quality factor $Q$ of the waveguide results in
$\ln Q \sim k\rho _0 (\epsilon _1/\epsilon )^{3/2}$. Since the
number $k \rho _0 = m $ is large, a relatively small fluctuation
of the dielectric constant within the area $2\pi \rho _0 w$ of the
ring can produce a large value of $Q$.

\begin{figure}
\centerline{
\epsfxsize=3.7in
\epsfbox{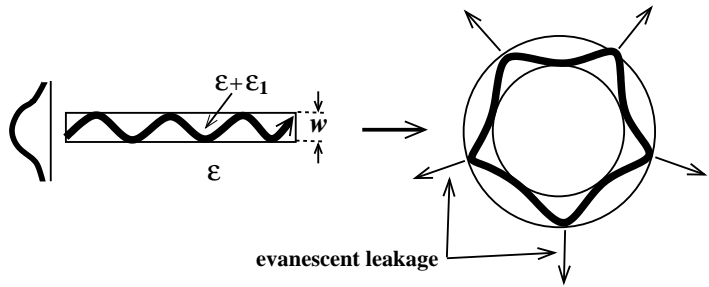}}
\vspace*{0.07in}
\caption[]{Rationale for the structure of the resonator.
Upon wrapping a stripe with enhanced dielectric constant
into a ring, a waveguided mode transforms into a
whispering-gallery mode.
}
\label{eps2}
\end{figure}

 The probability ${\cal W}$ for creating the required fluctuation strongly
depends on $R_c$. For a short range disorder ($k_0 R_c \ll 1$)
fluctuations of order $\epsilon _1$ should occur {\em
independently} in a large number, $N \sim \rho _0 w /R_c^2$, of
spots within the ring, so that the probability ${\cal W} \sim \exp
\left( - N \epsilon _1^2  /\Delta ^2 \right)$. In the other
extreme of strongly correlated disorder, when $R_c \gg  w$, the
number of independent spots is much smaller, $N ^{\prime } \sim
\rho _0 /R_c $ (the number of squares with a size $R_c $ needed to
cover the ring). Correspondingly, the probability ${\cal W} \sim
\exp \left( - N^{\prime } \epsilon _1^2/\Delta^2 \right)$ is much
larger than for the short range case. Finally, using the relation
between $\epsilon _1 $ and $Q$, the probability ${\cal W} = \exp (
-S_m) $ can be rewritten in terms of $Q$, thus, yielding an
estimate for $S_m$. For the short range case ($k_0 R_c \ll 1$) we
obtain $S_m \sim kl \ln Q$, where the mean free path $l$ is
proportional to $(R_c \Delta )^{-2}$ [see Eq. (\ref{kl4})]. In the
opposite limit of a smooth disorder  Eq. (\ref{kl4}) yields for
the transport mean free path  $l \sim R_c /\Delta ^2$.
 Then we have $S_m \sim N^{\prime } \epsilon _1^2/\Delta^2 \sim
l (\ln Q )^{4/3}/(k R_c^2 m^{1/3})$. Thus, for given $kl$ and $Q$,
the density of resonators is the higher the smoother is the
disorder. This conclusion is central to our study and will be
addressed below in more detail.

\subsubsection{Quantitative Results from the Optimal Fluctuation Approach.}

The above program can be carried out analytically \cite{apalkov02} with the use
of the optimal fluctuation approach \cite{Halperin66,Zittartz66}. This approach
is based on the idea that, when the exponent, $S_m $, in $\mbox{\bf  N}_m(kl,Q)$
is large, then the major contribution to $\mbox{\bf  N}_m(kl,Q)$
comes  from a certain specific disorder realization. In application to random
resonators,  the optimal fluctuation procedure
reduces to finding the {\em most probable} fluctuation of the dielectric constant
 which is able to trap the light for a long time $\sim \omega^{-1}Q$.
Assuming that the fluctuation is azimuthally symmetric (see Fig. 2), the shape
of the optimal fluctuation can be found explicitly \cite{apalkov02}
yielding
the following expression for exponent $S_m$
\begin{equation}
 S_m \! =2^4 3^{-3/2} \pi ^{1/2} m
\left(\frac{\epsilon _1^3}{\epsilon  }\right)^{1/2}
 \!  \frac{\Phi (\epsilon _1^{1/2}k_0R_c )}
{(\Delta  k_0 R_c)^2} ~,
   \label{eq7}
\end{equation}
where $\epsilon _1 = \epsilon (3 \ln Q/2m)^{2/3}$.
The analytical expression for  the function $\Phi (u)$ is the following
\begin{equation}
\Phi(u)  \! = \!
\frac{3^{3/2}}{2^{6}}
\frac{(5+ \sqrt{9+16 u^2})^{5/2}}
{(3+ \sqrt{9+16 u^2})^{3/2}} .
  \label{eq8}
\end{equation}
Recall now, that we are interested in the density of random
resonators at a {\em given } value of $kl$. The transition from
$\Delta $ to $l$ is  accomplished by using Eq.~(\ref{kl4}). For
the short range case, when $R_c \rightarrow 0$ and
 $\Phi (u) \rightarrow 1$, we obtain
\begin{equation}
S_m (k_0 R_c \ll 1)
   =  2\left( \frac{\pi ^3 }{3 } \right)^{1/2} kl  \ln Q
              ~. \label{eq2}
\end{equation}
To trace the change of $S_m$ with increasing $R_c$ it is convenient, after
using Eq.~(\ref{kl4}), to present Eq.~(\ref{eq7}) in the form
\begin{equation}
\frac{ S_m(k_0R_c > 1)}{ S_m(k_0R_c \ll  1)} =
\frac{ \Phi(\epsilon _1^{1/2} k_0 R_c)}
{\pi ^{1/2} (\epsilon ^{1/2} k_0 R_c)^3}~.   \label{eq1}
\end{equation}
 It is seen from Eq.~(\ref{eq1})
that $S_m$ falls off rapidly with increasing $R_c$.
In the domain $k_0 R_c > 1 $, but $\epsilon _1 ^{1/2} k_0 R_c
< 1$ we have $\Phi \approx 1$, so that
 $S_m \propto (k_0 R_c)^{-3}$. For larger $R_c$
we have $\Phi (u) \propto u $. In this domain
$S_m$ decreases slower with $R_c $: $S_m \propto (k_0 R_c)^{-2}$
\begin{equation}
\label{smooth}
S_m (k_0 R_c \gg 1)= \frac{3^{4/3}\pi}{4^{5/3}}\frac{kl\ln^{4/3} Q}{m^{1/3}(kR_c)^2}.
\end{equation}
Asymptotic expressions (\ref{eq2}) and  ({\ref{smooth}) agree with
the results of the qualitative derivation, with all numerical
factors now being determined. We emphasize that Eqs.~(\ref{eq2})
and (\ref{smooth}) apply {\em for a given} $kl$ value, so that the
decrease of $S_m$ with $R_c$ {\em leaves the backscattering cone
unchanged}.

\subsubsection{Estimates.}

Equation (\ref{eq2}) quantifies the effectiveness of trapping of
light in a random medium with point-like scatterers.  It follows
from Eq.~(\ref{eq2}) that the likelihood of high-$Q$ cavity is
really small. Indeed, even for rather strong disorder, $kl = 5$,
the exponent, $S_m$, in the probability of having a cavity with a
quality factor $Q=50$ is close to $S_m = 120$.  We emphasize that
in two dimensional case under consideration, this exponent does
not depend on $m$ and, thus, on the cavity radius $\rho _0 =
m/\epsilon ^{1/2} k_0$. More accurate
calculation \cite{apalkov'02}, taking into account the corrections
to Eq.~(\ref{eq2}), indicates that $S_m$ as a function of $m$ has a
minimum at  $m \sim \left(kl \ln Q\right)^{1/2}$.

To estimate the degree to which finite size of scatterers ($\sim
R_c$) improves the situation, we choose $k_0 R_c \approx 2$, which
already corresponds to the limit $k_0 R_c \gg 1$ in Eq.~(\ref{kl4}),
but still allows to set $\Phi = 1$. Then for $Q=50$, $kl = 5$ we
obtain $S_m \approx 1.1$, suggesting that the resonators with this $Q$
are quite frequent. In the latter estimate we have set $\epsilon =4$.

\subsection{Frequently Asked Questions}

\subsubsection{Why Rings?}

The answer to this question is illustrated in Fig. 4. The
distinguishing property of a ring is that the local curvature
radius is the same at each point. Upon any deviation from the ring
geometry, the curvature in a  certain region of the fluctuation
would be higher than in all other regions. Since the evanescent
losses are governed by this curvature, the quality factor of the
resonator would be  determined exclusively by this region (see
Fig. 4), so that the remaining low-curvature part would be
``unnecessary'', in the sense, that a ring with a radius
corresponding to the maximal curvature would have the same
quality factor as a square in Fig.~4 but significantly higher
probability of formation.
It is also quite obvious that, for the purpose of
supporting a wave-guided mode of the whispering-gallery type, a
ring is much superior to a disk of the same radius: indeed, the
internal area of the disk remains unused in the guiding process,
whereas a heavy penalty in terms of probability is paid in
creating this area.

\begin{figure}
\centerline{
\epsfxsize=3.3in
\epsfbox{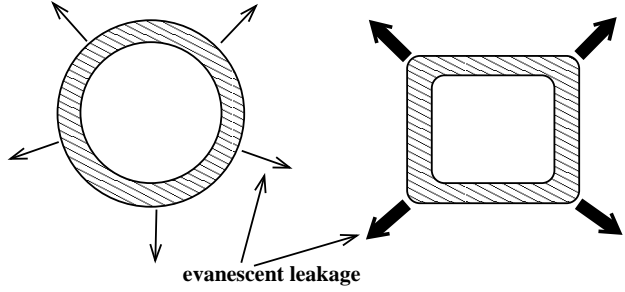}}
\vspace*{0.07in}
\caption[]{This drawing illustrates the optimal character of
the ring-shape resonator. }
\label{eps4}
\end{figure}

\subsubsection{Why Smooth Disorder Facilitates Trapping?}

At the qualitative level, the enhancement of the probability of
formation of the  cavity with increasing $R_c$ can be understood
for a toy model of the disorder, illustrated in Fig.~5. Suppose
that all the disks, that model the scatterers, are identical. Then
$R_c$ scales with the radius of the disk, $R$. Since the disks
cannot interpenetrate, the ring-shaped cavity corresponds to their
arrangement in the form of a necklace. The probability of
formation of such a cavity can be estimated as follows. Suppose
that a  sector, $\delta \phi$, is ``allocated'' for a single disk.
The probability to find a disk within this sector, at the distance
$\rho_0$ from the center , is $\sim n (\rho _0 \delta \phi )^2$,
where $n$ is the concentration of the disks. Thus, the probability
of formation of the necklace is
$\exp \left[ -\frac{2\pi }{\delta \phi }\ln\left(\frac{1}{n \rho _0^2
  (\delta \phi )^2}\right) \right]$,
where $\frac{2\pi }{\delta \phi }$ is the number of sectors. It is
obvious that if a  necklace is ``loose'', the quality factor of
the corresponding cavity would be low. In order for $Q$ to be
high,  neighboring disks must almost touch each other. This
implies that $\delta \phi \approx 2R /\rho _0$. Then the above
estimate for probability takes the form $\exp \left[ -\frac{\pi
\rho_0}{R}\ln\left(\frac{1}{f}\right) \right]$, where $f = n \pi
R^2$ is the filling fraction.
 This probability increases exponentially with $R$, i.e., with $R_c$, reflecting
the fact that, for a given $\rho_0$, the number of disks to be arranged is smaller
when $R$ is larger.
The above estimate was based on the assumption that the positions
of the disks are uncorrelated, i.e., $f \ll 1$ (in contrast to
 \cite{vanneste01} where $f=0.4$).
We have used the model of {\em hard} disks
as an easiest illustration of the role of $R_c$.
Obviously, Eq.~(\ref{eq7}) does not
apply to this model, since it was derived under the assumption that
the  statistics of the fluctuations is gaussian.

\begin{figure}
\centerline{
\epsfxsize=3.3in
\epsfbox{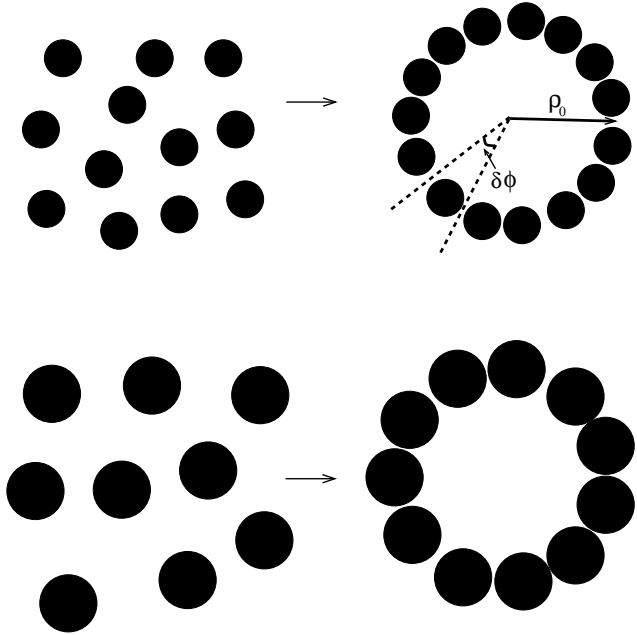}}
\vspace*{0.07in}
\caption[]{ A schematic illustration explaining why larger correlation
radius for a fixed filling fraction facilitates trapping.
A sector, shown with dashed lines, illustrates the tolerance
in the arrangement of disks into a necklace. }
\label{eps5}
\end{figure}

\subsubsection{``Vulnerability'' of the Ring-Shaped Cavities.}

The value $S_m$ given by Eq.~(\ref{eq7}), which was derived within the
optimal fluctuation approach, is the exponent in the probability
of formation of an {\em ideal} ring. Obviously,
any actual disorder realization is not ideal, in the sense,
that actual distribution of dielectric constant differs from the
optimal. For the same reason, the probability of formation of
ideal necklace of the type shown in Fig.~5 is zero.
In order for the probability to be finite, we should allow a certain
{\em tolerance} in the positions of the centers of the disks, as
illustrated in Fig.~5. In the conventional applications of the
optimal fluctuation approach \cite{Brezin80}, deviations from the
optimal distribution do not affect the value of the exponent, $S_m$.
However, in application to the random cavities, we
have searched for the fluctuation which is optimal {\em for
a given trapping time}, $\omega ^{-1} Q$. In this particular
application, a ``normal'' gaussian deviation from the optimal geometry can have a
catastrophic effect on trapping by scattering the light wave out
of the whispering-gallery trajectory. This scattering is discussed below.

{\em Scattering within the plane}. Two-dimensional picture adopted
throughout this paper, implies that electromagnetic field is
confined within a thin film in the $z$-direction. This confinement
results from the fact that the average dielectric constant of the
film is higher than in the adjacent regions. Then the filed
distribution,   ${\cal E}_0(z)$, along $z$ corresponds to a
transverse waveguide mode. For a given frequency, $\omega$, the
almost localized state on a ring can be destroyed due to the
scattering into states with the {\em same} distribution of the
field in the $z$-direction, which propagate freely along the film.
More precisely, the almost localized state with a given angular
momentum $m$, which is protected from the outside world by the
centrifugal barrier, can be scattered out to the continua of
states with smaller $m$'s, for which there is no barrier. It is
essential to estimate the lifetime, $\tau$, with respect to these
scattering processes and to verify that it is feasible to have
$\tau$ larger than the prescribed trapping time, $\omega ^{-1} Q$,
so that the almost localized state is not destroyed. A rigorous
treatment of this "scattering out" effect is quite involved and is
done in Section III, for a Gaussian potential.

The effect can be also illustrated with the
model of randomly positioned   hard disks (Fig.~5), although the
disorder in this model is non-gaussian. It is seen from  Fig.~5,
that  spacings between the rings, which are due to tolerance, open
a channel for the light escape, that is different from evanescent
leakage. A {\em typical}  lifetime with respect to such an escape
is quite short, i.e., even a small tolerance, which affects weakly
the exponent in the probability of the cavity formation, seems to
be detrimental for trapping. At this point we emphasize that, in
calculation of the scattering rate out of the whispering-gallery
trajectory, the disks constituting the necklace must be considered
as a {\em single entity}. As a result, for a {\em given}
configuration of the disks, the rate of scattering out caused by a
{\em single} disk
 must be
multiplied by the following {\em form-factor}
\begin{equation}\label{form1}
{\cal F}= \int \frac{ d\varphi _{\bbox{k}}}{2\pi } \left| \sum _{i}
\exp \left(i \bbox{k} \bbox{\rho }_i \right) \right|^2 =
\sum_{i,j}J_0\left(k\vert \bbox{\rho}_i - \bbox{\rho}_j\vert\right),
\end{equation}
where $\bbox{\rho}_i$ is the position of the center of the $i$-th disk in the
necklace. The form-factor, ${\cal F}$, is the sum of
$\left(\frac{\pi\rho_0}{R}\right)^2\gg 1$ terms. Out of this number,
$\left(kR\right)^{-1}\left(\frac{\pi\rho_0}{R}\right)$ terms (for $kR < 1$) and
$\frac{\pi\rho_0}{R}$ terms (for $kR > 1$),
for which $k\vert \bbox{\rho}_i - \bbox{\rho}_j\vert < 1$, are close to unity.
The portion of these terms is small. Other terms have random sign. This leads us
to the important conclusion that, for certain realizations of the necklace in Fig.~5,
the form-factor can take anomalously small values. For these realizations the quality
factor will be still determined by the evanescent leakage. The ``phase volume'' of
these realizaions is exponentially small and depends strongly on the  model
of the disorder.

{\em  Scattering out of the plane}. Compared to the previous case, two modifications
are in order. Firstly, since the final state of the scattered cavity mode
is a plane wave with the wave vector pointing in a certain direction within the
solid angle $4\pi$, the expression Eq. (\ref{form1}) for the form-factor should
be replaced by
\begin{equation}\label{form2}
\tilde{\cal F}=
\sum_{i,j}\frac{\sin\left(k_0\vert \bbox{\rho}_i - \bbox{\rho}_j\vert\right)}
{k_0\vert \bbox{\rho}_i - \bbox{\rho}_j\vert}.
\end{equation}
Secondly, for  $kR >1$, i.e., when the disorder is smooth,
scattering out of the plane that is caused by a single disk,
requires a large wave vector
transfer, $\sim k$. Thus, the corresponding rate is suppressed as compared to
the in-plane scattering.

\section{Prefactor}

\subsection{Qualitative discussion}

In Sec.~II we calculated the {\em probability} of
the formation of a trap that captures light for
a long time $\tau = \omega ^{-1} Q$. However, the relevant
characteristics of traps is their {\em areal density}.
Within the optimal fluctuation approach,
the relation between the probability and the areal
density emerges in course of calculation of the
prefactor \cite{Brezin80,North}.
Namely, the combination with dimensionality of the inverse
area comes from the so called ``zero modes'', which reflect
the fact that the fluctuation can be shifted
{\em as a whole} in both $x$ and $y$ directions.
 A typical shift, making two fluctuations independent, is
of the order of
the extent of the fluctuation in each direction.
This suggests that the proportionality coefficient between
the density of traps and the probability of the formation of
the trap is roughly  the inverse area of the fluctuation \cite{Brezin80}.
In our particular case, when the fluctuation is ring-shaped, the estimate
for the {\em dimensional} prefactor is $\sim w^{-2}$,
where $w$ is the width of the ring (see Sec.~II).
The dimensionless part of the prefactor within the
optimal fluctuation approach reflects the
``phase volume'' of the fluctuations, which  perturb
the shape of the optimal fluctuation leaving
the ``energy'' $k^2$ unchanged.

For the almost localized modes, considered above, the situation
with prefactor is qualitatively different from the case of the
truly localized states, for which the optimal fluctuation approach
was devised \cite{Halperin66}. The specifics of the almost
localized states is that their ``energy'', $k^2$, [see
Eq.~(\ref{Eq1})] is degenerate with continuum of the propagating
modes. As a result, a typical small perturbation of the shape of
the ring will not only shift $k^2$, but also cause the coupling of
the trapped mode to the continuum, or in other words, the {\em
additional leakage} will emerge due to the fluctuations. In order
to incorporate this effect into the theory, the density of traps
should be multiplied by
the probability, $P(\tau )$, that the lifetime with respect
to this additional
leakage is longer than $\tau $ \cite{apalkov'02}.
To estimate this probability, we consider the additonal leakage
for a {\em given} disorder realization
\begin{equation}
\frac{1}{\tau _{\mbox{\tiny $U$}}} = 2 \pi \left(\frac{c}{k}\right)
          \sum _{\mu }  \left|
  \int d \bbox{\rho } ~ \psi _0^{*} (\bbox{\rho }) ~\! U(\bbox{\rho })
     ~ \! \psi _{\mu } (\bbox{\rho })
    \right|^2  \delta (k_{\mu }^2 - k^2) ,
\label{golden_0}
\end{equation}
where $\psi _0$ is the ``localized'' solution of Eq.~(\ref{Eq1})
(without evanescent leakage), $\psi _{\mu }$ and $k_{\mu }$
are the propagating eigenfunctions and eigenvalues of Eq.~(\ref{Eq1}),
respectively.
It is seen from Eq.~(\ref{golden_0}) that small additional
leakage
requires small matrix elements
$\langle \psi_0 | U | \psi _{\mu } \rangle $.
Since the functions $\psi _{\mu }$ belong to the continuous spectrum,
it might seem
that this requirement is impossible to meet. This is, however,
not the case, since normalization factor of $|\psi _{\mu }|^2$ is the
inverse system size. To demonstrate that additional leakage can be
small,
it is convenient to use the explicit form of $\psi _{\mu }$, namely,
the plane waves, and rewrite Eq.~(\ref{golden_0})  in the form
\begin{equation}
\frac{1}{\tau _{\mbox{\tiny $U$}}} = \int \!\!\! \int d \bbox{\rho }_1
~ \! \! d \bbox{\rho }_2
      U(\bbox{\rho }_1)
     \mbox{\large $S$}(\bbox{\rho }_1,\bbox{\rho }_2)
U( \bbox{\rho }_2) ,
\label{golden_2}
\end{equation}
where the kernel $S(\bbox{\rho }_1,\bbox{\rho }_2)$ is defined as
\begin{eqnarray}
\mbox{\large $S$}(\bbox{\rho }_1,\bbox{\rho }_2) & = &
 \psi_0^* (\bbox{\rho }_1 ) \psi _0 (\bbox{\rho }_2)
                            \left(\frac{c}{k}\right)
\int \frac{d\bbox{q}}{2\pi}
    e^{i\bbox{q} (\bbox{\rho }_1 -\bbox{\rho }_2)} \delta (q^2 - k^2)   \nonumber \\
    & = &
 \frac{c}{2k} \psi_0^* (\bbox{\rho }_1 ) \psi _0 (\bbox{\rho }_2) J_{0}
  \left(k |\bbox{\rho }_1  - \bbox{\rho }_2 | \right) ,
\label{S1}
\end{eqnarray}
It is  seen from Eq.~(\ref{S1}) that the kernel
$\mbox{\large $S$}(\bbox{\rho }_1,\bbox{\rho }_2)$ is
exponentially small if one of the points, $\bbox{\rho }_1$ or $\bbox{\rho }_2$,
is located outside
the region, occupied by the ``body'' of $\psi _0 (\bbox{\rho }) $.
 When both $\bbox{\rho }_1$ and $\bbox{\rho }_2$ are located
inside this region, then the characteristic spatial scale of
the kernel, $\mbox{\large $S$}$, is $\left| \bbox{\rho }_1 - \bbox{\rho }_2 \right|
\sim k ^{-1}$.
Thus, for the sake of our qualitative discussion, we can
replace $\mbox{\large $S$}(\bbox{\rho }_1,\bbox{\rho }_2)$ by
$A^{-1}\theta \left(  k ^{-1} - \left| \bbox{\rho }_1 - \bbox{\rho }_2 \right|  \right)$,
where $A\sim |\psi _0(0)|^{-2}$ is the area of the fluctuation,
and restrict integration in Eq.~(\ref{golden_2}) to the region of the area $A$.

Averaging over disorder configurations in Eq.~(\ref{golden_2}) yields
the mean value of the additional leakage
\begin{equation}
\frac{1}{\tau _e} = \left(\frac{c}{k}\right)
     \int  \frac{d\bbox{q}}{2\pi} \left\langle \left|
  \int d \bbox{\rho } ~ \psi _0^{*} (\bbox{\rho }) ~\! U(\bbox{\rho })
     ~\! e^{i\bbox{q} \bbox{\rho }}
    \right|^2 \right\rangle \delta (q^2 - k^2) = \frac{c}{l_e},
\label{golden_1}
\end{equation}
which is determined by {\em typical} values of $U(\bbox{\rho })$.
We note that $l_e$ is of the order of the mean free path, $l$.
The fact that $l_e \sim l$ can
be seen from Eqs.~(\ref{golden_2})-(\ref{golden_1}),
taking into account that the area $A$ is always bigger than $R_c^2$.
Moreover, for short-range disorder, Eqs.~(\ref{golden_2})-(\ref{golden_1})
suggest that $l_e \approx l$.

In order to have $\tau _{\mbox{\tiny $U$}}^{-1}$ small, the
actual value of $U(\bbox{\rho })$ should be suppressed with respect to
typical in {\em each} box of a size $k^{-1}$. Then the condition
$\tau _{\mbox{\tiny $U$}} > \tau $ requires the suppression factor
to exceed $\tau /\tau _e$. The corresponding probability can be
estimated as
\begin{equation}
P(\tau )=(\tau _e/\tau )^{N}=(Q/\omega \tau_e )^{-N},
\label{P0}
\end{equation}
where $N\sim k^2 A$ is the number of boxes
that ``cover'' the area of localization of $\psi _0$. In particular,
for the ring-shaped fluctuations we have $N\sim w \rho _0 k^2$.
We conclude that, due to additional leakage, the prefactor in the
density of traps is {\em exponentially} small, i.e.,
$P(\tau )\sim \exp [-  k^2 A \ln (Q/kl)]$. The exponent $k^2 A \ln Q$
should be
compared with the main exponent, found in Sec.~II.
Since the above estimate assumed a short-range disorder, $kR_c \ll 1$,
the principal exponent is
$2\left(\pi ^3 /3 \right)^{1/2} kl  \ln Q$ [Eq.~(\ref{eq2})]. Recall that,
for short-range disorder we have $w \sim k^{-1} (k\rho _0/\ln Q )^{1/3}$,
$\rho _0 \sim m/k$, so
that the product $k^2 A$ is $\sim m^{4/3} \ln ^{-1/3} (Q/kl)$. The final
estimate for the exponent, originating from the prefactor, is
$\sim m^{4/3} \ln ^{2/3} Q$. This suggests that traps are the more frequent
the smaller is the angular momentum, $m$. On the other hand, in
derivation
of the main exponent we have assumed that $m > \ln Q$. So that the
minimal value of $ (m^{2} \ln  Q)^{2/3}$ is $ \sim \ln^2  Q$.
This leads us to the conclusion that the main exponent {\em dominates}
over the exponent, originating from the prefactor. This is because for
the main exponent we have $kl \ln Q > \ln ^2 Q$, since the relation
$kl > \ln Q$ was assumed in the derivation of the main exponent.

The above  qualitative analysis leading to Eq.~(\ref{P0})
was restricted, for
simplicity, to the case of the short-range disorder.
In the next subsection we present a rigorous calculation
of $P(\tau )$ for the more realistic case of a smooth disorder.

\subsection{Derivation of $P(\tau )$  }

Denote by $p(\tau )$ the probability {\em density } that the lifetime
with respect to additional leakage is equal to $\tau $, so that
$p(\tau ) = dP(\tau )/d \tau $. The rigorous definition of this density
reads
\begin{equation}
p(\tau ) = {\cal N}  \int {\cal D}\{ U \}
~ e^{-{\cal P}\{ U\}   }~
\mbox{\large $\delta$} \!
\left( \tau  - \tau_{\mbox{\tiny $U$}} \right) ,
\label{P2_1}
\end{equation}
where the normalization constant is defined as
\begin{equation}
{\cal N} =
\left[  \int {\cal D}\{ U \} ~e^{-{\cal P}\{ U\}}  \right]^{-1}
\label{norm}
\end{equation}
and ${\cal P} \{ U \}$ is given by
\begin{equation}
{\cal P} \{ U \} = \frac{1}{2 U_0^2} \int \!\!\! \int d\bbox{\rho }_1 d\bbox{\rho }_2
              U(\bbox{\rho }_1)
\mbox{\large $\kappa$} (\bbox{\rho }_1,\bbox{\rho }_2 ) U (\bbox{\rho }_2),
\label{calP1}
\end{equation}
where $\mbox{\large $\kappa$} (\bbox{\rho }_1,\bbox{\rho }_2 )$ is related to the
correlator ${\mbox{\large $K$}(\bbox{\rho_1},  \bbox{\rho_2})}$,
defined by Eq.~(\ref{correlator}),
as
\begin{equation}
\int  d\bbox{\rho }^{\prime }
             \mbox{\large $\kappa$} (\bbox{\rho }_1,\bbox{\rho }^{\prime } )
{\mbox{\large $K$}(\bbox{\rho}^{\prime },  \bbox{\rho }_2)} =
 \delta \left(\bbox{\rho }_1-\bbox{\rho }_2  \right).
\label{inverse}
\end{equation}
Since we are dealing with photons, the value $U_0$ can be expressed
through the r.m.s. fluctuation of the dielectric constant as
$U_0 = k_0^2 \Delta $ [see Eq.~(\ref{light})].
Using the integral representation of the $\delta $-function,  Eq.~(\ref{P2_1})
 can be rewritten in the form
\begin{equation}
p(\tau ) = \frac{{\cal N}}{2\pi }
       \int_{-\infty }^{\infty } dt ~ e^{it/\tau } \int {\cal D}\{ U \}
     ~ e^{-{\cal P}_t \{ U\}  }  ,
\label{P2_2}
\end{equation}
where the auxilary functional, ${\cal P}_t \{ U\}$, is defined as
\begin{equation}
{\cal P }_{t} \{ U\}= {\cal P} \{ U \} + i\frac{t}{\tau _{\mbox{\tiny $U$}}} =
 \int\!\!\! \int d \bbox{\rho }_1 ~ \!\! d \bbox{\rho }_2  U(\bbox{\rho }_1)
     \mbox{\large ${\cal K}$}_t (\bbox{\rho }_1,\bbox{\rho }_2) U( \bbox{\rho }_2).
\label{Pfunctional}
\end{equation}
The kernel, $\mbox{\large $ {\cal K}$}_t (\bbox{\rho }_1,\bbox{\rho }_2)$,
of the functional ${\cal P }_{t} \{ U\}$ has the form
\begin{equation}
    \mbox{\large $ {\cal K}$}_t (\bbox{\rho }_1,\bbox{\rho }_2)  =
  \frac{1}{2 }
    k_0^{-4} \Delta ^{-2}
   \mbox{\large $\kappa $} (\bbox{\rho }_1,\bbox{\rho }_2 )
     +  it \mbox{\large $S$}(\bbox{\rho }_1,\bbox{\rho }_2)
\label{K_total}
\end{equation}
Following the standard procedure of functional integration,
we present the fluctuation $U(\bbox{\rho })$ as a linear combination
\begin{equation}
U(\bbox{\rho }) = \sum _{\mu } C_{t, \mu } \phi _{t,\mu } ( \bbox{\rho })
\label{UC}
\end{equation}
where $ \phi _{t,\mu } ( \bbox{\rho })$ are the eigenfunctions of the
operator $\mbox{\large $\hat{{\cal K}}$}_t$ with the kernel
$\mbox{\large ${\cal K}$}_t $
\begin{equation}
 \mbox{\large $\hat{{\cal K}}$}_t  \phi _{t,\mu }  (\bbox{\rho }) =
    \int d\bbox{\rho }_1 \mbox{\large ${\cal K}$}_t (\bbox{\rho },\bbox{\rho }_1)
       \phi _{t,\mu } (\bbox{\rho }_1)=
     \Lambda _{t, \mu } \phi _{t,\mu } ( \bbox{\rho }),
\label{eigen}
\end{equation}
and $\Lambda _{t, \mu }$ are the corresponding eigenvalues.

At this point, we note that the operator $\mbox{\large $\hat{{\cal K}}$}_t$ is
non-hermician. As a consequence, the functions $\phi _{t,\mu }$ form an
orthogonal basis if the definition of the scalar product is modified to
$\left\langle \phi_1 | \phi_2 \right\rangle = \int d\bbox{\rho }~\!
\phi_1 (\bbox{\rho } ) ~ \! \phi _2 (\bbox{\rho }) $. To see this, note that
the kernel $\mbox{\large ${\cal K}$}_t
(\bbox{\rho }_1,\bbox{\rho }_2)$ is symmetric with respect to interchange
$\bbox{\rho }_1 \! \leftrightarrow \! \bbox{\rho }_2$,
$\mbox{\large ${\cal K}$}_t (\bbox{\rho }_1,\bbox{\rho }_2) = \mbox{\large ${\cal K}$}_t
(\bbox{\rho }_2,\bbox{\rho }_1)$. Then it follows from Eq.~(\ref{eigen})
\begin{eqnarray}
& & \int \!\!\! \int d\bbox{\rho }_1 d\bbox{\rho }_2  ~\!
     \phi _{t,\mu _1} (\bbox{\rho } _1 )   ~\!
 \mbox{\large ${\cal K}$}_t (\bbox{\rho }_1,\bbox{\rho }_2) ~\!
     \phi _{t, \mu _2} (\bbox{\rho }_2)    =
       \nonumber \\
&  &~~~~~~~~~~~~~~~~~~~~~~~~~~ =
        \int d\bbox{\rho }_1 ~\! \! \phi _{t, \mu _1} (\bbox{\rho } _1 )
                                 \!  \int  d\bbox{\rho }_2 ~\!
 \mbox{\large ${\cal K}$}_t (\bbox{\rho }_1,\bbox{\rho }_2) ~\!
     \phi _{t, \mu _2} (\bbox{\rho }_2)  \nonumber  \\
     &  &  ~~~~~~~~~~~~~~~~~~~~~~~~~~
  =  \Lambda _{t ,\mu _2}  \int d\bbox{\rho }_1
\phi_{t, \mu _1} (\bbox{\rho }_1 ) ~ \! \phi _{t, \mu _2} (\bbox{\rho }_1) .
\label{orth1}
\end{eqnarray}
On the other hand, it also follows from Eq.~(\ref{eigen})
\begin{eqnarray}
& & \int \!\!\! \int d\bbox{\rho }_1 d\bbox{\rho }_2  ~\!
     \phi _{t,\mu _1} (\bbox{\rho } _1 )   ~\!
 \mbox{\large ${\cal K}$}_t (\bbox{\rho }_1,\bbox{\rho }_2) ~\!
     \phi _{t, \mu _2} (\bbox{\rho }_2)
      =  \nonumber \\
 & & ~~~~~~~~~~~~~~~~~~~~~~~~~~
 = \int d\bbox{\rho }_2 ~\!  \phi _{t, \mu _2} (\bbox{\rho } _2 )  \!
        \int  d\bbox{\rho }_1 ~\! \!
 \mbox{\large ${\cal K}$}_t (\bbox{\rho }_2,\bbox{\rho }_1) ~\!
     \phi _{t, \mu _1} (\bbox{\rho }_1)  \nonumber  \\
     &  &  ~~~~~~~~~~~~~~~~~~~~~~~~~~  =
  \Lambda _{t ,\mu _1}  \int d\bbox{\rho }_2   ~\!
\phi_{t, \mu _1} (\bbox{\rho }_2 ) ~ \! \phi _{t, \mu _2} (\bbox{\rho }_2).
\label{orth2}
\end{eqnarray}
Comparing the r.h.s. of Eqs.~(\ref{orth1}) and (\ref{orth2}), we have
\begin{equation}
\left\langle \phi_{t, \mu _1} | \phi_{t, \mu _2} \right\rangle  =
\int d\bbox{\rho } ~\!
\phi_{t, \mu _1} (\bbox{\rho } ) ~\! \phi _{t, \mu _2} (\bbox{\rho })
 = \delta _{\mu _1, \mu _2}.
 \label{orth3}
\end{equation}
Using the expansion Eq.~(\ref{UC}), the functional integral Eq.~(\ref{P2_2})
reduces to  the integration over {\em all} complex coefficients $C_{t, \mu }$
in the expansion Eq.~(\ref{UC}). However, due to the restriction
that $U(\bbox{\rho })$ is real, the pair $\{ \mbox{Re}C_{t, \mu },
\mbox{Im}C_{t, \mu }\}$ can be ``rotated'' in such a way that
instead of integrals over real and imaginary parts we get a single integral
along the real axis.
\begin{equation}
p(\tau ) = \frac{1}{2\pi }
   \int_{ -\infty }^{\infty } dt~ e^{i t/\tau }~
    \frac{ \int _{ -\infty }^{\infty } {\cal D}\left\{ \tilde{C}_{t, \mu }\right\}
\exp \left( - \sum _{\mu } \Lambda _{t, \mu } \tilde{C}_{t, \mu }^2 \right)
         }{
\int _{ -\infty }^{\infty }{\cal D}\left\{ \tilde{C}_{0, \mu }\right\}
\exp \left( - \sum _{\mu } \Lambda _{0, \mu }  \tilde{C}_{0, \mu }^2 \right)
         }   .
\label{P2_4}
\end{equation}
To proceed further, we note that the real parts of all the eigenvalues
$\Lambda _{t,\mu }$ are positive. Indeed, it follows from Eqs.~(\ref{K_total}) and
(\ref{eigen}) that
\begin{equation}
\mbox{Re} \Lambda _{t,\mu } =
  \left( \frac{1}{2
    k_0 ^4 \Delta ^{2}
  } \right)
\frac{
\int \!\!\! \int d\bbox{\rho }_1 d\bbox{\rho }_2
              \phi_{t,\mu }^{*} (\bbox{\rho }_1)
\mbox{\large $\kappa$} (\bbox{\rho }_1,\bbox{\rho }_2 ) \phi_{t,\mu } (\bbox{\rho }_2)}
{\int \!\!\! \int d\bbox{\rho } |\phi_{t,\mu } (\bbox{\rho } ) |^2 }   .
\end{equation}
Since $\mbox{\large $\kappa$} $ is positively defined, $\mbox{Re} \Lambda _{t,\mu }$
is positive.
The last remark ensures the convergence of all the gaussian integrals
in Eq.~(\ref{P2_4}). Thus, we obtain from Eq.~(\ref{P2_4})
\begin{equation}
p(\tau ) = \frac{1}{2\pi }
           \int_{ -\infty }^{\infty } dt ~e^{i t/\tau }
        \left( \prod_{\mu } \frac{\Lambda _{0,\mu }}{\Lambda _{t,\mu }}
     \right)^{1/2}  .
\label{P2_5}
\end{equation}
Since the product of the eigenvalues of an operator is equal to its
determinant, Eq.~(\ref{P2_5}) can be rewritten in the form
\begin{equation}
p(\tau ) =   \frac{1}{2\pi }
           \int_{ -\infty }^{\infty } dt ~
e^ {i t/\tau }~\!
        \left( \frac{\det \mbox{\large $\hat{{\cal K}}$}_0  }
                    { \det \mbox{\large $\hat{{\cal K}}$}_t  }   \right)^{1/2}.
\label{P2_6}
\end{equation}
It is convenient to present the ratio of determinants in
 the integrand of Eq.~(\ref{P2_6}) as a single determinant. This is achieved
through the following sequence of steps
\begin{eqnarray}
 \frac{\det \mbox{\large $\hat{{\cal K}}$}_0  }
      { \det \mbox{\large $\hat{{\cal K}}$}_t  }
 & = &
  \frac{1}{ \det \mbox{\large $\hat{{\cal K}}$}_0^{-1}
        \det \mbox{\large $\hat{{\cal K}}$}_t  }
  =
  \frac{1}{ \det \left(  \mbox{\large $\hat{{\cal K}}$}_0^{-1}
    \mbox{\large $\hat{{\cal K}}$}_t \right) }
  =  \nonumber \\
  & = & \frac{1}{  \det \left[ 2 k_0^4 \Delta ^2
   \mbox{\large $\hat{\kappa }$} ^{-1} \! \left( \frac{1}{2} k_0^{-4} \Delta ^{-2}
\mbox{\large $\hat{\kappa }$}   +  it \mbox{\large $\hat{S }$}   \right)  \right]     }
   =  \nonumber \\
  & = & \frac{1}{  \det \left[  1  + 2 it k_0^4 \Delta ^2  ~\!
  \mbox{\large $\hat{ K}$}  \mbox{\large $\hat{S }$}  \right]     }  ,
\label{det2}
\end{eqnarray}
where we have used the explicit form (\ref{K_total}) of the operator
$ \mbox{\large $\hat{{\cal K}}$}_t $. The operators
 $\mbox{\large $\hat{ K}$}$ and $\mbox{\large $\hat{S }$}$
are the integral operators with the kernels
 $\mbox{\large $S$}(\bbox{\rho }_1,\bbox{\rho }_2)$ [Eq.~(\ref{S1})] and
  $\mbox{\large $K$}(\bbox{\rho }_1,\bbox{\rho }_2)$, respectively.
We recall that the operator $\mbox{\large $\hat{ K}$}$
 is the inverse of the operator  $\mbox{\large $\hat{ \kappa }$}$
[Eq.~(\ref{inverse})].
Upon transformation (\ref{det2}), the expression (\ref{P2_6})
for $p(\tau )$ takes the form
\begin{eqnarray}
p(\tau ) & = &  \frac{1}{2\pi }
           \int_{ -\infty }^{\infty } dt
       \frac{e^ {i t/\tau }}{  \sqrt{ \det \left[  1  +
                         2 it k_0^4 \Delta ^2  ~ \!
\mbox{\large $\hat{K}$} \mbox{\large $\hat{S}$}    \right] } }
   \nonumber \\
 & = &  \frac{1}{\pi }
        \int_{0}^{\infty } dt ~\mbox{Re}\left\{
       \frac{e^ {i t/\tau }}{  \sqrt{ \det \left[  1  + 2 it k_0^4 \Delta ^2
\mbox{\large $\hat{K}$} \mbox{\large $\hat{S}$}
                                              \right] }     }\right\}  .
\label{P2_7}
\end{eqnarray}
To proceed further, we need to analyze the properties of eigenfunctions and
eigenvalues of the operators $\mbox{\large $\hat{ K}$}$ and
$\mbox{\large $\hat{S }$}$.

\subsubsection{Properties of operator $\mbox{ $\hat{ K}$}$.}

It is easy to see that the  eigenfunctions
of $\mbox{\large $\hat{ K}$}$ are plane waves. Indeed, since the
kernel $\mbox{\large $K$}$ depends only on the
difference $(\bbox{\rho }_1 - \bbox{\rho }_2)$, we have
\begin{equation}
\mbox{\large $\hat{ K}$}e^{i \bbox{p}\bbox{\rho }} =
\int d \bbox{\rho _1} \mbox{\large $K$}(\bbox{\rho } - \bbox{\rho }_1)
e^{i \bbox{p}\bbox{\rho }_1}=
\mbox{\large $\tilde{K}$} (p) e^{i \bbox{p}\bbox{\rho }} ,
\end{equation}
so that eigenvalues of $\mbox{\large $\hat{ K}$}$ are the
Fourier components of the correlator $\mbox{\large $K$}(\rho )$.
Thus, these eigenvalues are strongly suppressed if  $p> R_c^{-1}$.
For the particular case of gaussian correlator we have
\begin{equation}
 \mbox{\large $\tilde{K}$} (p)    =
                      \int d \bbox{\rho } ~\!\! \mbox{\large $K$}(\rho )
 ~\! e^{i \bbox{p}\bbox{\rho } }
 =  \pi R_c^2  e^{-p^2R_c^2/4}  .
 \label{Kp}
\end{equation}

\subsubsection{Properties of operator $\mbox{ $\hat{ S}$}$.}

Although the eigenfunctions, $\xi _{\mu }(\bbox{\rho })$,
of the operator $\mbox{\large $\hat{ S}$}$ are not plane
waves, their width in the $k$-space is narrow (of the
order of inverse spatial extent of the function $\psi _0$) as
it can be seen from Eq.~(\ref{S1}). To estimate the
eigenvalues, $\lambda _{\mu }$, of $\mbox{\large $\hat{ S}$}$,
defined as
\begin{equation}
\int d \bbox{\rho _1} \mbox{\large $S$}(\bbox{\rho },\bbox{\rho }_1)
\xi _{\mu }(\bbox{\rho }_1) =
  \lambda _{\mu }  \xi _{\mu }(\bbox{\rho }) ,
\end{equation}
we first note that these eigenvalues satisfy the following
sum rule
\begin{equation}
\sum _{\mu }  \lambda _{\mu }
  = \frac{c}{2k}  .
\label{sum2}
\end{equation}
This rule follows from the identity
\begin{eqnarray}
\sum _{\mu }  \lambda _{\mu }  & = &
\int \!\!\!
\int  d \bbox{\rho }~ \! d \bbox{\rho _1}
                      \mbox{\large $S$}(\bbox{\rho },\bbox{\rho }_1)
 \sum _{\mu } \xi ^{*} _{\mu } (\bbox{\rho})
                                 \xi _{\mu }(\bbox{\rho }_1)  \nonumber \\
 & = &
  \int  d \bbox{\rho } ~\mbox{\large $S$}(\bbox{\rho },\bbox{\rho })
=
\frac{c}{2k}
    \int  d \bbox{\rho } ~|\psi_0 (\bbox{\rho } )|^2 ,
\label{sum1}
\end{eqnarray}
in which the completeness of the set $\xi _{\mu }(\bbox{\rho })$ is used,
so that
\begin{equation}
 \sum _{\mu } \xi ^{*} _{\mu } (\bbox{\rho}) \xi _{\mu }(\bbox{\rho }_1)  =
\delta ( \bbox{\rho } - \bbox{\rho }_1  ).
\end{equation}
The reason why Eq.~(\ref{sum2}) allows to estimate the eigenvalues
is their specific distribution.
 Namely, the first $N \approx k^2 A$
eigenvalues are almost equal to each other, while the eigenvalues
with numbers  $\mu > N$ fall off rapidly, faster than $(N/\mu)^4$.
This rapid fall off  of $\lambda _{\mu }$ has a simple explanation.
The eigenfunctions corresponding to large $\mu $ are close to
plane waves, so that the value of $\lambda _{\mu }$ is determined
by the integral of the rapidly oscillating plane wave over the
area $A$ and is small due to the cancellation. On the contrary,
for $\mu < N $ the eigenfunction changes weakly within the
area $A$. Thus, for $\mu < N$ we have $\lambda _{\mu } \approx
\lambda _0 \sim c/(kN)$. For the particular case of the ring-shaped
fluctuations $N \approx 2\pi \rho _0 w k^2 $, so that
$\lambda _0 \approx c/(\rho _0 w k^3)$.

\subsubsection{Evaluation of the integral (42).}
{\em Short-range disorder.}
 In this case $\mbox{\large $\tilde{K}$}(p)\approx
\mbox{\large $\tilde{K}$}(0) \equiv \pi R_c^2 $,
so that the contribution to the integrand of Eq.~(\ref{P2_7})
comes from $N$ eigenvalues of the operator
$\mbox{\large $\hat{S }$}$, i.e.,
 $\det \left[  1  + 2 it k_0^4 \Delta ^2
\mbox{\large $\hat{K}$} \mbox{\large $\hat{S}$}   \right]
 \approx \left[  1  + 2 i\pi t k_0^4 \tilde{\Delta} ^2 \lambda _{0} \right]^{N}$,
 where $R_c \Delta \rightarrow \tilde{\Delta}$ when $R_c \rightarrow 0$.
Then the integration over $t$ in (\ref{P2_7}) can be easily performed,
yielding
\begin{eqnarray}
p(\tau ) & = & \frac{1
}{4 \tau (N /2-1)!}
  \left( 2 \pi ~\! \tau \lambda _{0 }~\! k_0^4 \tilde{\Delta }^2  \right)^{2-N/2}
\exp \left[ - \left( 2 \pi  \tau \lambda _{0 } k_0^4 \tilde{\Delta}^2 \right)^{-1}
        \right]
  \nonumber \\
& \sim  &
 \exp\left[-\frac{N}{2} \ln\left(
\frac{Q}{kl_e}
\right) -
   \left(\frac{N}{8}\right) \frac{kl_e}{Q}\right]    ,
\label{sr}
\end{eqnarray}
where we have used the large-$N$ asymptotics of $N!$ and
the fact that $\lambda _0 \approx c/(kN)$.
Since the first term in the exponent is much larger than the
second one, we recover with exponential accuracy the form of $P(\tau )$,
obtained in the qualitative consideration.

\vspace{3mm}

{\em Smooth disorder.} In the case of a smooth disorder, $k R_c >1$,
it is the fast decay of $ \mbox{\large $\tilde{K}$}(p )$ [Eq.~(\ref{Kp})]
rather than $\lambda _{\mu }$,
that introduces a ``cutoff'' of the determinant in Eq.~(\ref{P2_7}).
In this case it is convenient to rewrite the determinant in Eq.~(\ref{P2_7})
in the form
\begin{eqnarray}
\det \left[  1  + 2 it k_0^4 \Delta ^2
\mbox{\large $\hat{K}$} \mbox{\large $\hat{S}$}  \right]
 &  = &
 \prod_{\mu } \left[  1  + 2 it \lambda _0 ~\! k_0^4 \Delta ^2
\mbox{\large $\tilde{K}$} (p_{\mu })  \right] \nonumber \\
 & = & \exp \left[ \sum_{\mu }\ln \left(1+ 2 it \lambda _0 ~ \! k_0^4 \Delta ^2
                         \mbox{\large $\tilde{K}$}(p_{\mu }) \right)
         \right]    .
\label{D0_2}
\end{eqnarray}
The sum over $\mu $ in the exponent of (\ref{D0_2}) goes over both
projections, $p_x$ and $p_y$, of the momentum $p$. For ring-shaped
fluctuations it is natural to consider the radial and azimutal components
of $p$. Since $w\sim R_c$ and $\rho _0 \gg R_c$ the contribution to the
sum comes only from a single radial component, while the sum over
angular component can be replaced by an integral. Thus we obtain
\begin{equation}
\det \left[  1  + 2 it k_0^4 \Delta ^2
\mbox{\large $\hat{K}$} \mbox{\large $\hat{S}$}  \right]
  = \exp \left[ \rho _0 \int _{0}^{\infty } dp ~
   \ln  \!  \left(1+ 2 it \lambda _0 ~\! k_0^4 \Delta ^2   \mbox{\large $\tilde{K}$}(p) \right)
         \right]   .
\label{D0_3}
\end{equation}
The main contribution to the integral (\ref{D0_3}) comes from the domain
$\tau \lambda _0 ~ \! k_0^4 \Delta ^2  \mbox{\large $\tilde{K}$}(p) \sim 1$, so that
$p > R_c^{-1}$. It is instructive to perform further calculations for more
general form of $\mbox{\large $\tilde{K}$}(p )$, namely
$\mbox{\large $\tilde{K}$}(p ) \sim \exp[ - (pR_c)^n ]$, which reduces to
Eq.~(\ref{Kp}) when $n=2$. Substituting this form into Eq.~(\ref{D0_3}) and
performing two subsequent integration by parts, we get
\begin{eqnarray}
 \det \left[  1  + 2 it k_0^4 \Delta ^2
\mbox{\large $\hat{K}$} \mbox{\large $\hat{S}$}  \right]
  & = & \exp \left[
  \frac{i \rho _0}{R_c}
  \left( \frac{n}{n+1}
  \right)
  \int _{0}^{\infty } dw ~
      w^{(n+1)/n}\frac{e^{w-w_t}}{ \left( e^{w-w_t} +i \right)^2} \right]  \nonumber \\
  & = & \exp \left[  \frac{i \rho _0}{R_c} \left(\frac{n}{n+1}\right)  w_t^{(n+1)/n}
  \int _{-\infty }^{\infty } dw ~ \frac{e^{w}}{ \left( e^{w} +i \right)^2} \right] ,
  \label{D0_4}
\end{eqnarray}
where  $w_t = \ln (2 \pi t \lambda _0 R_c^2 k_0^4 \Delta ^2 )$. In the second identity
we made use of the fact that the function
$\frac{e^{w-w_t}}{ \left( e^{w-w_t} +i \right)^2}$ has a sharp
maximum at $w = w_t$. The remaining integral in  Eq.~(\ref{D0_4}) can be evaluated
exactly, $\int _{-\infty }^{\infty } dw ~
\frac{e^{w}}{ \left( e^{w} +i \right)^2} = -i$, so that Eq.~(\ref{D0_4})
takes the form
 \begin{equation}
 \det \left[  1  + 2 it k_0^4 \Delta ^2
\mbox{\large $\hat{K}$} \mbox{\large $\hat{S}$}  \right]
  = \exp \left[
 \frac{\rho _0}{R_c} \left( \frac{n }{n+1 } \right)
  w_t^{(n+1)/n}\right]   .
\label{D0_5}
\end{equation}
Substituting this form into Eq.~(\ref{P2_7}), we obtain
\begin{equation}
 \! \! p(\tau ) = \frac{1}{\pi }  \!
       \int_{0}^{\infty } \! \! dt ~\! \cos (t/\tau )
   \exp \left[ -
 \frac{n \rho _0 }{2(n+1) R_c}
 \ln^{(n+1)/n}       \mbox{\Large$($} 2 \pi t \lambda _0 R_c^2 k_0^4 \Delta ^2
                  \mbox{\Large$)$}\right] .
\end{equation}
Evaluating the above integral with an exponential accuracy yields
\begin{equation}
 p(\tau ) \sim
   \exp \left[ -
 \frac{n\rho _0}{2(n+1)R_c} \ln^{(n+1)/n}
                  \mbox{\Large$($} 2 \pi \tau \lambda _0 R_c^2 k_0^4 \Delta ^2
                           \mbox{\Large$)$}
    \right]             .
\end{equation}
Since $p(\tau )$ and $P(\tau )$ have the same exponential dependence, the
final expression for the prefactor, $P(\tau )$, takes the form
\begin{equation}
P(\tau )  \sim
    \exp \left[ -
  \frac{n\rho _0}{2(n+1)R_c} \ln^{(n+1)/n}\left( Q/kl_a \right)
                  \right]      ,
\label{P_final}
\end{equation}
where $\tau $ in the argument of the logarithm was replaced by the trapping
time $\omega ^{-1} Q$. For particular case of the gaussian correlator
(\ref{Kp}) the probability that the lifetime with respect to additional
leakage is longer than $\omega ^{-1} Q$ is given by
\begin{equation}
 P(\omega ^{-1} Q )  \sim
    \exp \left[ -
  \frac{\rho _0}{3 R_c} \ln^{3/2}(Q /kl_a)
                  \right]      .
\label{P_gauss}
\end{equation}
Note, that for the short-range disorder, [Eq.~(\ref{sr})],
the number $N$ was
the number of sections with the area $\sim k^{-2}$, which ``cover''
the ring-shaped trap. Correspondingly, for the smooth disorder the ratio
$\rho _0/R_c$ in the exponent is the number of squares with the side
$\sim R_c$ that cover the trap. On the other hand, as it is
seen from Eq.~(\ref{P_final}), the  power of the logarithm is specific
for the gaussian correlator.

\subsection{Optimal Ring}

Combining the main exponent for the gaussian correlator, Eq.~(\ref{smooth}), and
the corresponding prefactor, Eq.~(\ref{P_gauss}), we obtain
\begin{eqnarray}
 \mbox{\bf N}_m(Q) & = & P(\omega _c Q) e^{-S _m (Q)}  \nonumber \\
    &\sim & \exp \left[ -
  \frac{m }{3k R_c} \ln^{3/2}Q
 -
  3^{4/3}4^{-5/3} \pi  \frac{kl}{(kR_c)^2}
  m^{-1/3} \ln^{4/3} Q  \right] .
\label{P_total_g}
\end{eqnarray}
It is seen that the dependence of the main exponent and
the prefactor on $m$ are opposite. The prefactor, reflecting the
``vulnarability'' of the ring, falls off with $m$, while the
main exponent favors large $m$ value for which the $Q$-factor is
higher. As a result of the competition between the two
tendencies $\mbox{\bf N}_m(Q)$ has a sharp maximum at optimal
value of $m$ given by
\begin{equation}
m_{opt} =
    \frac{3 \pi ^{4/3}}{4^{5/4}} \left(\frac{l}{R_c}  \right)^{3/4}
    \ln ^{-1/8} Q
\end{equation}
Substituting $m=m_{opt}$ into Eq.~(\ref{P_total_g}), we
arrive at the final result
\begin{equation}
\mbox{\bf N} _{opt} (Q) = \exp \left[ - 2^{-1/2} \pi ^{3/4}
\frac{(kl)^{3/4}}{(kR_c)^{7/4}}
  \ln ^{11/8} Q
\right]   .
\label{P_total_g_1}
\end{equation}
If we use a general form of the prefactor Eq.~(\ref{P_final}),
then the changes in Eq.~(\ref{P_total_g_1}) amount to
an additional factor $2^{-13/4}(3n)^{1/4}(n+1)^{-5/4}(11n+2)$.
Also the power of $\ln Q$ in  Eq.~(\ref{P_total_g_1}) modifies to
$(5n+1)/4n$. Overall, these changes are inessential, so that
the result, Eq.~(\ref{P_total_g_1}), is rather robust. It shows how
the trapping is enhanced due to a smooth disorder, when the
additional leakage is taken into account. Without the prefactor
this enhancement manifested itself through the combination
$(k R_c)^{2}$ in the denominator of the main exponent, $S_m$.
With the prefactor, $(k R_c)^{2}$ is replaced by $(kR_c)^{7/4}$, so that
the enhancement is weaker, but insignificantly.

As a final remark, we note that additional leakage, caused by the scattering
out of the plane, can be incorporated into the theory in a similar
fashion as the in-plane additional leakage. Corresponding changes are
outlined in the end of Sec.II. Recall also, that for the smooth
disorder the suppression of $\mbox{\bf N}_m (Q)$ due to additional
leakage is dominated by the in-plane scattering processes.

\section{Conclusion}
In the present paper we studied a new type of
solutions \cite{apalkov02,karpov93} of the wave equation, Eq.
(\ref{Eq1}), in a weakly disordered medium. The solutions, dubbed
as "almost localized" states, describe a wave which is confined
primarily to a small ring. In an {\em open} sample, of size $L$
much smaller than the two-dimensional localization length $\xi$,
the almost localized states correspond to sharp resonances,
residing in the high-$Q$ ring-shaped cavities, as discussed
throughout the paper. However, in a {\em closed} sample -which
would require perfectly reflecting walls- the resonances turn into
true eigenstates, whose almost entire weight is located at the
rings. In this respect the "almost localized" states differ from
the "prelocalized" states, extensively studied in the context of
electronic transport [72-75].

  We have provided a quantitative
theory of the almost localized states and the associated random
resonators, and pointed out their relevance for the phenomenon of
random lasing. We stress, however, that  these random resonators
exist already in the {\em passive} medium, and gain is only needed
"to make them visible". Moreover, the resonators are
"self-formed", in the sense that no sharp features (like Mie
scatterers or other "resonant entities") are introduced: the model
is defined by Eq.(\ref{Eq1}), which describes a featureless
dielectric medium with fluctuating dielectric constant.

 Our
study substantiates the intuitive image  \cite{cao98,cao99,cao99''}
of a resonant cavity as a closed-loop trajectory of a light wave
bouncing between the point-like scatterers. The intuitive picture
in  \cite{cao98,cao99,cao99''} assumed that light can propagate
along a loop of scatterers by simply being scattered from one
scatterer to another. Such a picture, however, is unrealistic due
to the scattering out of the loop.  We have demonstrated that the
scenario of light traveling along closed loops can be remedied. In
our picture the "loops", i.e., the random resonators, can be
envisaged as rings with dielectric constant larger than the
average value.
The reason why such rings are able to trap the light is that
the constituting scatterers act as a {\em single entity}:
only the coherent multiple scattering of light by {\em all}
the scatterers in the resonator can provide trapping.
We have also established that correlations in the
fluctuating part of the dielectric constant
highly facilitate trapping.

We acknowledge the support of the National Science Foundation
under Grant No. DMR-0202790 and of the Petroleum Research
Fund under Grant No. 37890-AC6.

\end{document}